\documentclass[a4paper]{jpconf}
\usepackage{graphicx}
\usepackage{lineno}
\begin{document}
\modulolinenumbers[2]
\linenumbers

\title{Measurements of $\mathrm{J/\psi \rightarrow e^{+}e^{-}}$ with ALICE at the LHC}

\author{Fiorella Fionda on behalf of the ALICE Collaboration}

\address{Dipartimento Interateneo di Fisica ``M. Merlin'' and Sezione INFN, Bari, Italy}

\ead{fiorella.fionda@ba.infn.it}

\begin{abstract}
The ALICE detector provides excellent capabilities to study 
quarkonium production at the Large Hadron Collider (LHC). 
Heavy quarkonia, bound states of charm or beauty quark 
anti-quark pairs such as the $\mathrm{J/\psi}$, are expected to be 
produced by initial hard processes. Thus they will provide 
insight into the earliest and hottest stages of AA 
collisions where the formation of a Quark-Gluon Plasma (QGP)
is expected. Furthermore, high-precision data from pp 
collisions represent an essential baseline for the 
measurement of nuclear modifications in heavy-ions and serve
also as a crucial test for several models of quarkonium 
hadroproduction. In addition, the study of pA collisions 
allows to investigate nuclear modifications due to Cold 
Nuclear Matter (CNM) effects.
In ALICE, $\mathrm{J/\psi}$ were measured in pp and Pb--Pb collisions down 
to $p_{\mathrm T}$ = 0 via their di-electron decay channel in the central 
barrel ($|y| <$ 0.8). Results on the nuclear modification factor 
($R_{\mathrm{AA}}$) at central rapidities in Pb--Pb collisions at 
$\sqrt{s_{\mathrm{NN}}} = 2.76$ TeV will be shown and their implications
 discussed. A separation of the prompt and non-prompt components is also 
possible down to $p_{\mathrm T}$ of the $\mathrm{J/\psi}$ of 2 GeV/$c$. 
%possible down to \mbox{$p_{\rm T}$(J/$\psi$) = 2 GeV/$c$}. 
%which allows to study the beauty hadron nuclear modification 
%factor down to almost zero $p_{\rm T}$.
\end{abstract}

\begin{figure}[h]
\centering
\begin{minipage}{15.0pc}
\includegraphics[width=15.0pc]{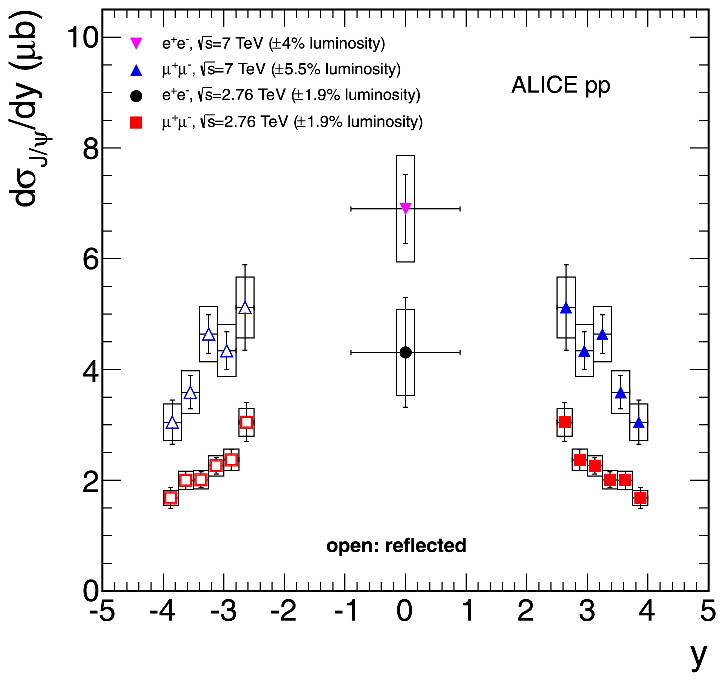}
\end{minipage}\hspace{2pc}%
\begin{minipage}{15.2pc}
\includegraphics[width=15.2pc]{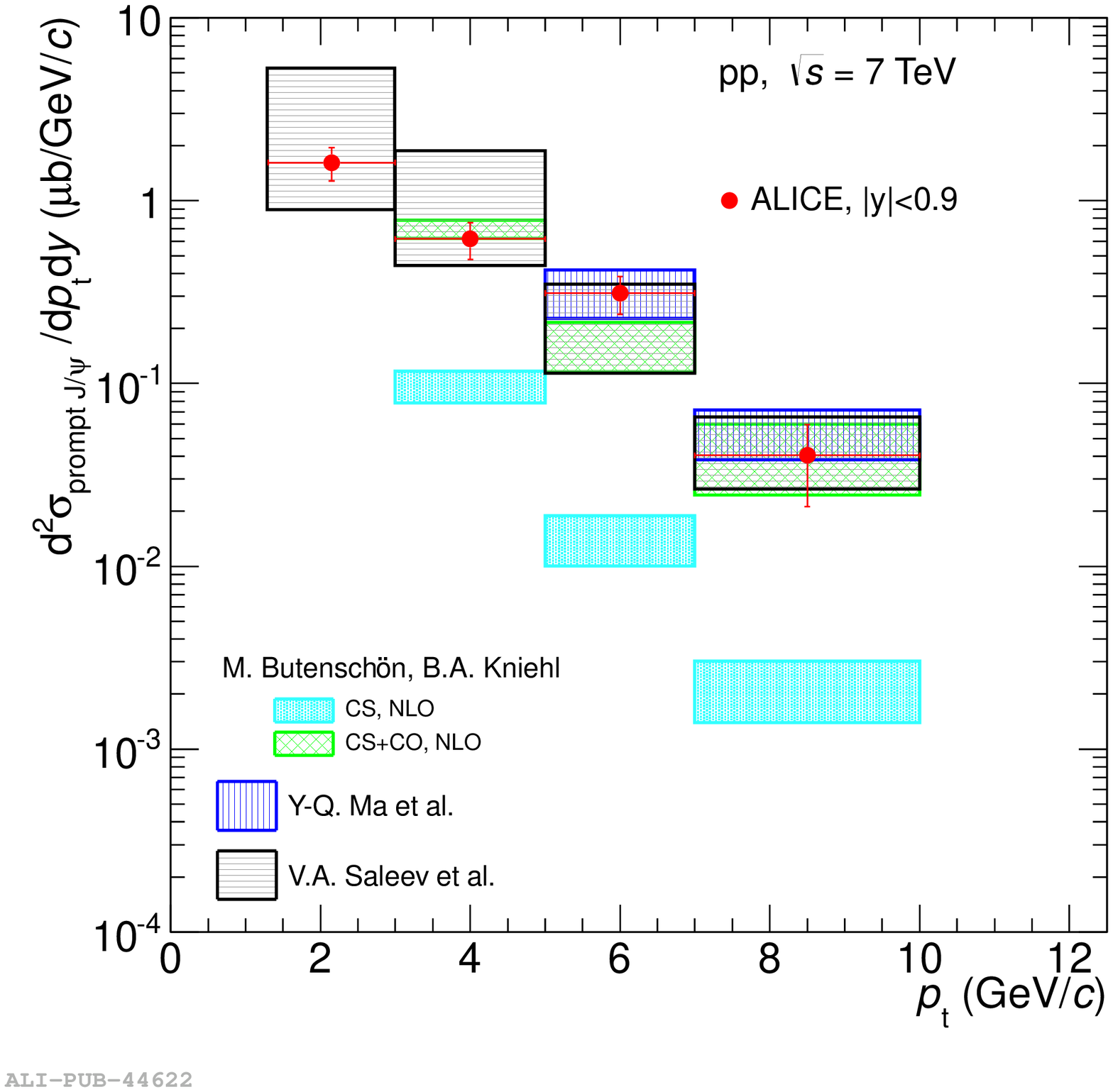}
\end{minipage}
\caption{\label{Xsec}Left: Inclusive $\mathrm{J/\psi}$ cross section as a function
of rapidity measured in pp collisions at $\sqrt{s} =$ 7 TeV and 2.76 TeV \cite{ALICEincl2.76TeV}. Right: Prompt $\mathrm{J/\psi}$ cross section as a function of 
$p_{\mathrm T}$ compared to several theoretical predictions \cite{ALICEnonPrompt7TeV}.}
\end{figure}

%\section{Physical motivations}
%{\bf Physical motivations.} 
The production of heavy quarkonia involves both perturbative
and non-perturbative mechanisms of Quantum-Chromo-Dynamics (QCD). 
In proton-proton collisions, several models \cite{mod1,mod2} 
attempted to describe 
the quarkonia production, but failed in reproducing simultaneously
cross-sections, polarization, transverse momentum and 
rapidity dependence as measured at the Tevatron \cite{Tev1,Tev2} and 
RHIC \cite{RHIC} colliders.
Results in proton-proton collision at the new LHC energies
have provided additional constraints to those models as well as
the baseline reference for AA analyses. 
At the high temperatures and large energy densities reached
in relativistic heavy-ion collisions, the matter 
consists of deconfined quarks and gluons, in the state 
referred to as ``Quark-Gluon-Plasma'' (QGP) \cite{QGP}.  
%Charmonia are very special probes to describe the properties
%of such a hot and dense medium.
According to the prediction by Matsui and Satz \cite{MatzuiSatz}, 
in the deconfined medium formed in
nucleus-nucleus collisions quarkonium production is
suppressed relative to that in proton-proton collisions due
to the color analogue of the Debye screening mechanism.
%According to the prediction by Matzui and Satz \cite{MatzuiSatz}
%that the production of quarkonia in nucleus-nucleus collisions 
%should be suppressed relative to that in proton-proton 
%due to the 
%Color-Debye Screening 
%color analogue of the Debye screening mechanism
%responsible for the melting of $\rm c\bar{c}$ states in
%the QGP. 
The observable to quantify the nuclear medium effects is the 
so-called ``nuclear modification factor'' $R_{\mathrm{AA}}$,
defined as 
%the ratio of the $\mathrm{J/\psi}$ production yield in nucleus-nucleus 
%collisions (AA) to that in pp, scaled by the number of binary 
%nucleon-nucleon collisions $\langle N_{\mathrm{coll}} \rangle$
%(estimated from Glauber model calculations):

\begin{eqnarray}
R_{\mathrm{AA}}  =  \frac{1}{\langle T_{\mathrm{AA}} \rangle} \frac{{\mathrm d^{2}}N_{\mathrm{J/\psi}}^{\mathrm{AA}}/{\mathrm d}p_{\mathrm T}{\mathrm d}y}{{\mathrm d^{2}}\sigma_{\mathrm{J/\psi}}^{\mathrm{pp}}/{\mathrm d}p_{\mathrm T}{\mathrm d}y} 
\nonumber
\end{eqnarray}

where $\langle T_{\mathrm{AA}} \rangle$ is the nuclear overlap
function determined by Glauber model calculations.
The $\mathrm{J/\psi}$ suppression observed 
at SPS and RHIC \cite{SPS,RHIC1,RHIC2} is not completely 
understood yet. Open points are in particular
%it has been often referred as ``anomalus suppression'' due to 
the observation of a similar $\mathrm{J/\psi}$ suppression at the two different 
centre-of-mass energies and a stronger
suppression at forward-rapidity ($\sim$40\%) compared
to mid-rapidity at RHIC.
Two theoretical models were proposed in order to reproduce 
RHIC and SPS data and provide predictions for the LHC: {\it i)} 
the ``regeneration'' mechanism from 
deconfined quarks in the medium to compete the $\mathrm{J/\psi}$ 
suppression in the QGP \cite{reg1,reg2}; {\it ii)} the
statistical hadronization of charm quarks at phase boundary 
\cite{statHadr1,statHadr2}.
%The interpretation of the experimental results is still under debate
%due to the lack of an exhaustive understanding of
%Cold Nuclear Matter (CNM) effects, 
%strongly dependent on the centre-of-mass energy of the system
%and determined from p-A collisions, 
%as well as the unknown total $\rm c\bar{c}$ cross section, which
%prevents more precise model calculations.
The interpretation of the experimental results is still under debate
due to the large experimental uncertainty on the total 
$\mathrm{c\bar{c}}$ production cross section, which
prevents more precise model calculations,
as well as the lack of an exhaustive understanding of
Cold Nuclear Matter (CNM) effects, 
strongly dependent on the centre-of-mass energy of the system
and determined from pA collisions.
Furthermore, other important contributions are $\mathrm{J/\psi}$ 
from the decays of higher mass charmonium 
states (e.g. $\mathrm{\chi_{\mathrm c}}$ and $\mathrm{\psi^{\mathrm \prime}}$)
and beauty hadrons (non-prompt $\mathrm{J/\psi}$). 
In particular, measuring the fraction of 
non-prompt $\mathrm{J/\psi}$, $f_{\mathrm B}$, gives access to
the $R_{\mathrm{AA}}$ of both prompt and non-prompt produced $\mathrm{J/\psi}$ mesons and
the latter reflects directly the nuclear modification factor
of beauty hadrons.
%which is stricktly connected to
%the b-quarks energy loss in the QGP. 
According to the
QCD predictions \cite{QCDpred} the parton energy
loss in the QGP implies the following hierarchy in the
measured $R_{\mathrm{AA}}$: $R_{\mathrm{AA}}^{\mathrm \pi} < R_{\mathrm{AA}}^{\mathrm D} < R_{\mathrm{AA}}^{\mathrm B}$. 
Therefore the comparison between the $R_{\mathrm{AA}}$ of non-prompt 
$\mathrm{J/\psi}$ with the $R_{\mathrm{AA}}$ of other hadrons 
could offer an important insight
onto the parton energy loss mechanisms in the QGP.
\\
\\
%\section{The ALICE detector at the LHC}
%{\bf The ALICE detector at the LHC.} 
%ALICE (A Large Ion Collider Experiment) is specifically designed for heavy-ion 
%physics, but it has also an important program with proton-proton collisions. 
%A complete description of ALICE detector can be found in the ref. \cite{ALICEdet}.
In ALICE the $\mathrm{J/\psi}$ production is measured at central rapidity 
($|y| < 0.8$) in the dielectron channel $\mathrm{J/\psi \rightarrow e^{+}e^{-}}$ 
and at forward rapidity ($2.5 < y < 4$) in the dimuon 
channel $\mathrm{J/\psi \rightarrow \mu^{+}\mu^{-}}$
reaching in both cases $p_{\mathrm T}$ = 0. 
The focus of this paper is on the results obtained at mid-rapidity.
%where
%the possibility to measure down to $p_{\rm T}$ = 0 is a unique
%feature at the LHC. 
The main tracking detectors used in this analysis are the Inner Tracking
System (ITS), which allows for the measurement of the
$\mathrm{J/\psi}$ fraction from beauty hadron decays, 
and the Time Projection Chamber (TPC), which is used for tracking 
and electron identification via specific energy deposition measurement. 
%The ITS is the vertex detector made up of 6 layers of silicon 
%detectors. Thanks to the two innermost layers 
%of Silicon Pixel detectors (SPD) the excellent impact 
%parameter resolution allows to measure the fraction of 
%J/$\psi$  ``displaced'' from primary vertex, i.e.
%coming from beauty hadrons decay.
%The TPC is the main tracking device and it is used
%also for electron identification, based
%on the measurement of their specific energy loss.
%Besides the TPC, the PID in central barrel (in particular for electrons) 
%is realized by adding the information of the Time Of Flight detector (TOF) 
%
%which helps to reject protons
%and pions for momenta below 1.5 GeV/$c$, and the 
%and the Transition Radiation Detector (TRD).
% which is dedicated to the electron-pion discrimination.
%
%\section{Results}
%The left-hand panel of Fig. \ref{Xsec} shows the inclusive J/$\psi$ cross section measured
%by ALICE in proton-proton collisions at $\sqrt{s}$ = 7 TeV as a function of transverse momentum
%and compared with similar results obtained by other experiments at the LHC 
%(see \cite{ALICEincl7TeV} and references therein). At central rapidity ALICE results 
%have a complementary $p_{\rm T}$ coverage w.r.t. the one accessible by ATLAS and CMS. 
\\
\\
%{\bf Results.} 
In the left-hand panel of Fig.~\ref{Xsec} the inclusive cross sections 
in pp collisions as a function
of rapidity are shown at both \mbox{$\sqrt{s}$ = 7 TeV} (\mbox{$L^{\mathrm{e^{+}e^{-}}}_{\mathrm{int}}$ = 5.6 nb$^{-1}$} and \mbox{$L^{\mathrm{\mu^{+}\mu^{-}}}_{\mathrm{int}}$ = 15.6 nb$^{-1}$}) 
and \mbox{$\sqrt{s}$ = 2.76 TeV} (\mbox{$L^{\mathrm{e^{+}e^{-}}}_{\mathrm{int}}$ = 1.1 nb$^{-1}$} and \mbox{$L^{\mathrm{\mu^{+}\mu^{-}}}_{\mathrm{int}}$ = 19.9 nb$^{-1}$}) \cite{ALICEincl2.76TeV}. 
The measurement at \mbox{$\sqrt{s}$ = 2.76 TeV} represents 
the reference used for the $R_{\mathrm{AA}}$ analysis in Pb--Pb.
The currently large uncertainties on this reference limit 
the accuracy on the $R_{\mathrm{AA}}$ determination.

The fraction of $\mathrm{J/\psi}$ coming from beauty hadron decays was 
measured in proton-proton collisions at 
$\sqrt{s}$ = 7 TeV down to $p_{\mathrm T}$ = 1.3 GeV/$c$. 
%The combination
%of this fraction with the inclusive J/$\psi$ cross sections lead
%to the determination of both prompt and non-prompt J/$\psi$ cross sections.
%from a 2-dimensional unbinned log-likelihood fit to the 
%$m_{inv}(e^{+}e^{-})$ and pseudoproper decay length\footnote{ 
%The pseudoproper decay length x is defined as 
%x  = $ \frac{L_{xy}m_{J/\psi}}{p_{T}^{J/\psi}}$
%where $L_{xy} = \frac{\vec{L}\cdot \vec{p}_{T}^{J/\psi}}{p_{T}^{J/\psi}}$
%is the vector from the primary vertex to the J/$\psi$ decay vertex.} distributions. 
In the right-hand panel of Fig.~\ref{Xsec}
the prompt $\mathrm{J/\psi}$ cross section as a function of transverse momentum
is compared to
%to NRQCD calculations
next-to-leading order (NLO) non-relativistic QCD (NRQCD) theoretical calculations 
%theoretical calculations by M. Butenschon and B.A. Kniehl 
%and Y.-Q. Ma et al. 
(see \cite{ALICEnonPrompt7TeV} and references therein),
%Both calculations 
which include color-singlet (CS) and color-octet (CO) contributions; heavier
charmonium feed-down is also included. The comparison suggests that the CO
processes are important to describe the data. 

\begin{figure}[h]
\centering
\begin{minipage}{17pc}
\includegraphics[width=17pc]{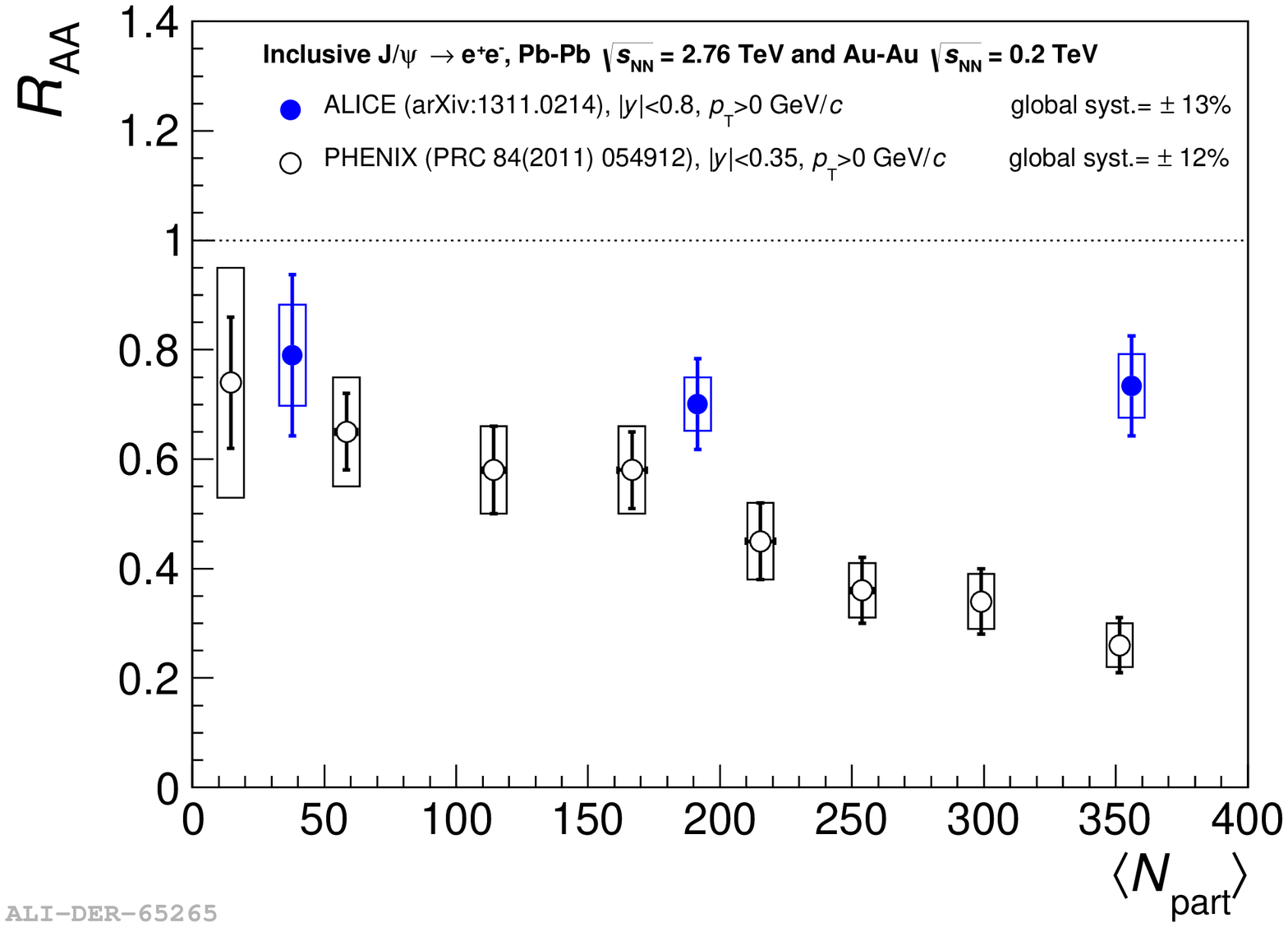}
\end{minipage}\hspace{2pc}%
\begin{minipage}{17pc}
\includegraphics[width=17pc]{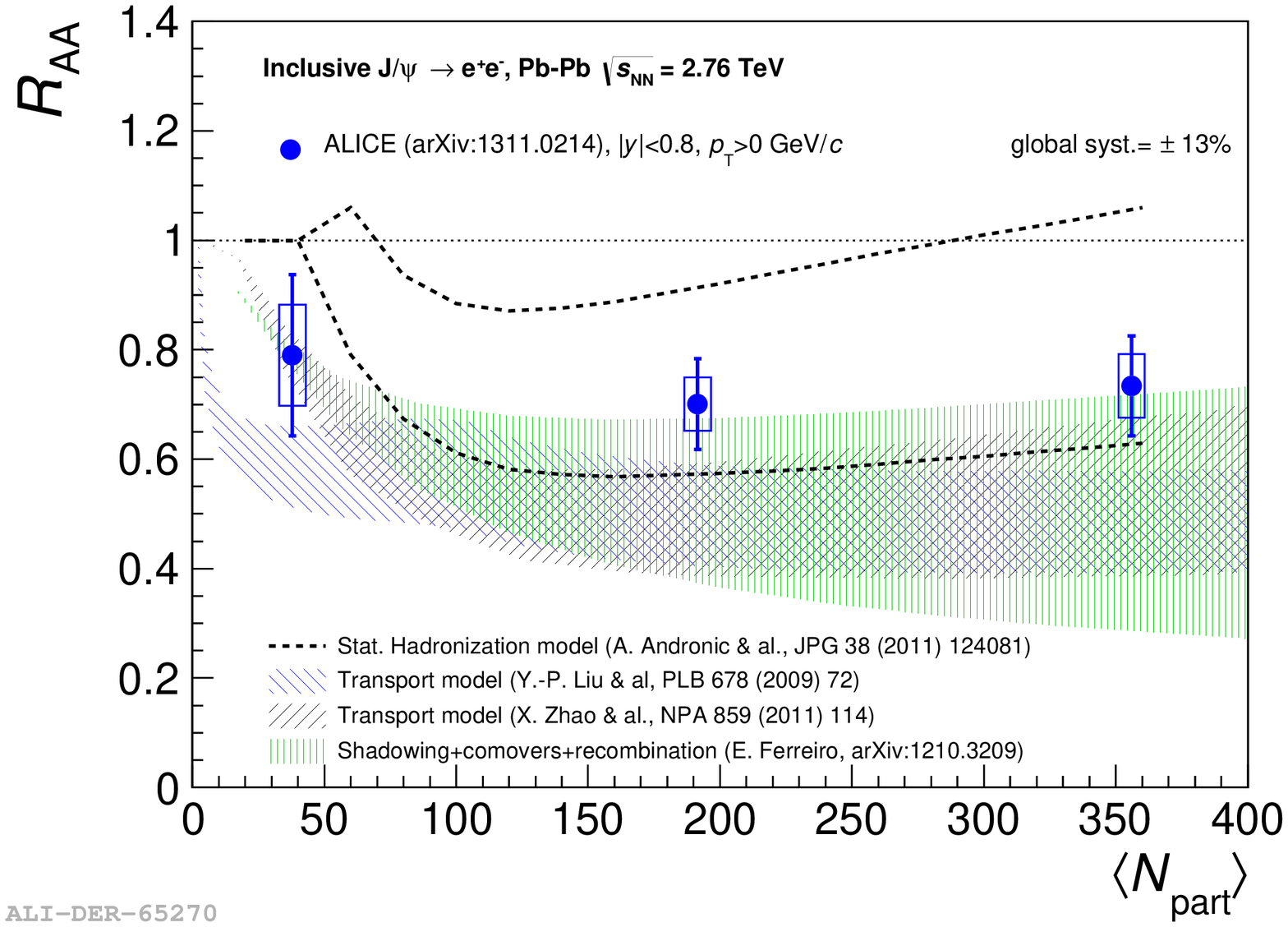}
\end{minipage}
\caption{\label{Raa} Nuclear modification factor $R_{\mathrm{AA}}$ measured
in Pb--Pb collisions at $\sqrt{s_{\mathrm{NN}}}$ = 2.76 TeV as a function
of the mean number of participants $\langle N_{\mathrm{part}} \rangle$ compared
with PHENIX \cite{Phenix} results at lower energy (left-hand panel) and with theoretical models \cite{reg1,reg2,comover,statHadr} (right-hand panel).}
\end{figure}

In Pb--Pb collisions at $\sqrt{s_{\mathrm{NN}}}$ = 2.76 TeV the nuclear modification factor 
$R_{\mathrm{AA}}$ of inclusive $\mathrm{J/\psi}$ was measured as a function of centrality 
for $p_{\mathrm T} >$ 0 ($L_{\mathrm{int}}$ = 15 $\mathrm{\mu}$b$^{-1}$). This is shown in the left-hand panel of Fig.~\ref{Raa} as a function of
the mean number of participant nucleons $\langle N_{\mathrm{part}} \rangle$ 
(estimated from Glauber model) and it
is compared with the inclusive $\mathrm{J/\psi}$ $R_{\mathrm{AA}}$ measured at mid rapidity
by PHENIX in Au-Au collisions at $\sqrt{s_{\mathrm{NN}}}$ = 200 GeV \cite{Phenix}.
The comparison indicates a reduced suppression for most central
collisions in ALICE w.r.t. PHENIX, and this behaviour is in qualitative
agreement with a regeneration scenario at LHC energies. 

In the right-hand panel of Fig.~\ref{Raa} the inclusive $\mathrm{J/\psi}$ $R_{\mathrm{AA}}$
is compared with theoretical models that include the
(re)combination of deconfined charm quark pairs from the QGP. 
In particular, the hashed bands represent 
the results from two transport models \cite{reg1, reg2} 
and from the comover interaction model \cite{comover} where up to 50\%
of the $\mathrm{J/\psi}$ were produced from 
deconfined ${\mathrm{c\bar{c}}}$ pairs recombination. 
The prediction from the statistical hadronization model \cite{statHadr},
shown by solid lines, is also shown. 
All models exhibit a good agreement with data albeit with large uncertainties, due to the large uncertainty on the  
inclusive ${\mathrm{c\bar{c}}}$ production cross section and 
CNM effects (e.g. nuclear shadowing). 
The latter are currently being addressed by measuring $\mathrm{J/\psi}$ 
production in p--Pb collisions.  

\begin{figure}[h]
\centering
\begin{minipage}{16.0pc}
\includegraphics[width=16.0pc]{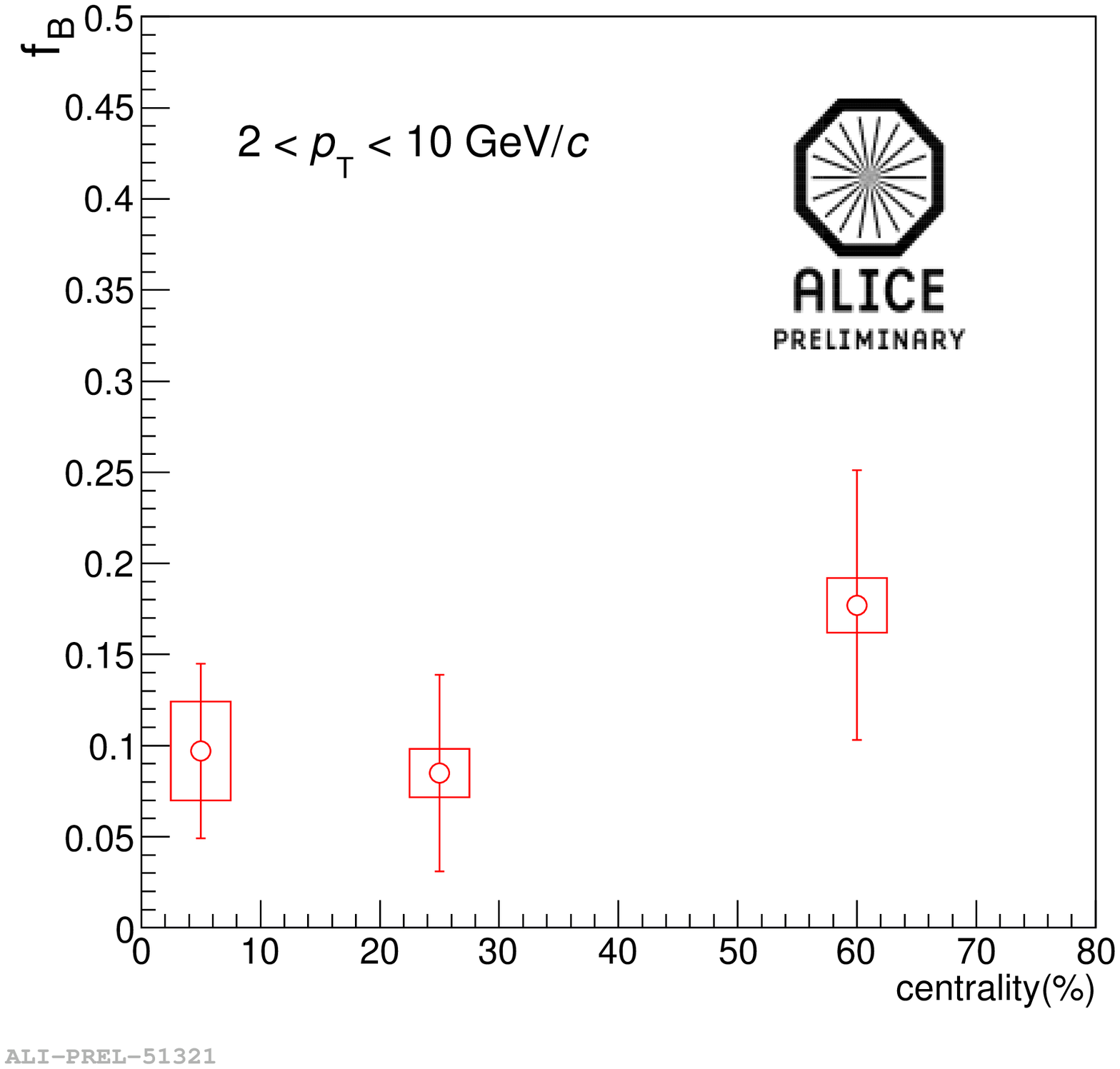}
\end{minipage}\hspace{2pc}%
\begin{minipage}{16.0pc}
\includegraphics[width=16.0pc]{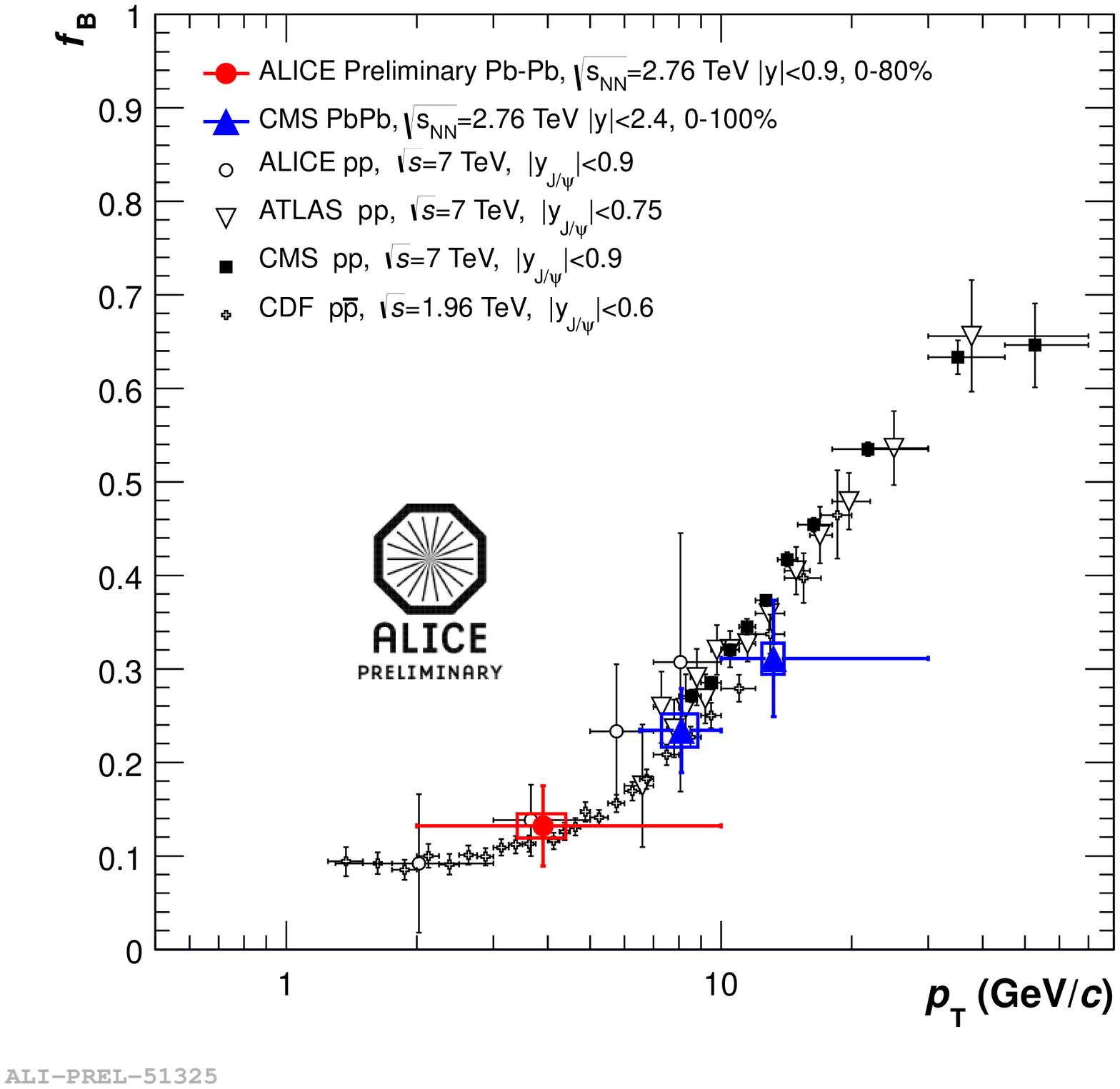}
\end{minipage}
\caption{\label{fBresults}Left: Non-prompt $ \mathrm{J/\psi}$ fraction measured in Pb--Pb collisions at $\sqrt{s_{\mathrm{NN}}}$ = 2.76 TeV as a function of centrality (statistical
and systematic uncertainties shown by bars and boxes respectively). Right: Fraction of non-prompt $\mathrm{J/\psi}$ in Pb--Pb collisions at $\sqrt{s_{\mathrm{NN}}}$ = 2.76 TeV measured by ALICE and CMS \cite{CMSPbPb} at central rapidity as a function of $p_{\mathrm T}$. Results in pp \cite{ALICEnonPrompt7TeV,ATLAS,CMS} and ${\mathrm{p\bar{p}}}$ \cite{Tev1} collisions are also shown.}
\end{figure}

The fraction of non-prompt $\mathrm{J/\psi}$ were measured in Pb--Pb collisions
at $\sqrt{s_{\mathrm{NN}}}$ = 2.76 TeV as a function of centrality down to $p_{\mathrm T}$ = 2 GeV/$c$,
as shown in the left-hand side of Fig.~\ref{fBresults}. 
%The error bars on the plot represent the statistical uncertainties
%whereas the boxes are the systematic uncertainties. 
No significant dependence
of the non-prompt $\mathrm{J/\psi}$ fraction on centrality can be observed. 
In the right-hand panel of Fig.~\ref{fBresults} the fraction of
non-prompt $\mathrm{J/\psi}$ as a function of transverse momentum 
measured by ALICE and CMS \cite{CMSPbPb} in Pb--Pb collisions and 
integrated over centrality is shown, along with the
results in pp collisions at $\sqrt{s}$ = 7 TeV
by ALICE \cite{ALICEnonPrompt7TeV}, ATLAS \cite{ATLAS} and 
CMS \cite{CMS}. The CDF data in ${\mathrm{p\bar{p}}}$ collisions at $\sqrt{s}$ = 1.96 TeV \cite{Tev1}
are also reported.
%to delineate the trend of $f_{B}$ in the lower $p_{T}$ region. 
Considering the ALICE and CMS results together, an indication for a similar
trend of $f_{\mathrm B}$ as a function of $p_{\mathrm T}$ in proton-proton and Pb--Pb collisions is observed.
\\
\\
%{\bf Conclusions.} 
In summary, transverse momentum spectra and rapidity distributions of inclusive $\mathrm{J/\psi}$
were measured down to $p_{\mathrm T}$ = 0 for pp collisions at $\sqrt{s}$ = 2.76 TeV and $\sqrt{s} =$ 7 TeV. NRQCD calculations are consistent with the measured prompt $\mathrm{J/\psi}$ production cross section.
%A good agreement for prompt $\mathrm{J/\psi}$ 
%cross section with NRQCD calculations is observed.   
The nuclear modification factor $R_{\mathrm{AA}}$ was measured in Pb--Pb 
collisions at $\sqrt{s_{\mathrm{NN}}}$ = 2.76 TeV down to \mbox{$p_{\mathrm T}$ = 0},
as a function of centrality. The comparison with 
PHENIX results and theoretical predictions provide an indication for
(re)generation of $\mathrm{J/\psi}$ from deconfined charm quarks. The non-prompt 
$\mathrm{J/\psi}$ fraction were also measured in Pb--Pb collision as
a function of centrality. The combination of ALICE and CMS
results in Pb--Pb collisions suggests a trend of $f_{\mathrm B}$ as a 
function of $p_{\mathrm T}$ that is similar to that in pp.

\end{document}